\documentclass[preprint,12pt]{elsarticle}
\usepackage{graphicx}
\usepackage{multicol}
\usepackage{color}
\usepackage{enumerate}
\usepackage{amssymb}
\usepackage{mathrsfs}
\usepackage{amsmath}
\usepackage{ulem}

\usepackage{amssymb}

\journal{xxx}

\begin{document}

\begin{frontmatter}



\title{Surface superconductor-insulator transition: Reduction of the critical electric field by Hartree-Fock potential}

\author[1]{Yajiang Chen\corref{cor1}}
\ead{yjchen@zstu.edu.cn}
\affiliation[1]{organization={Key Laboratory of Optical Field Manipulation of Zhejiang Province, Department of Physics, Zhejiang Sci-Tech University},
            postcode={320018}, 
            state={Zhejiang},
            country={China}}
            
\cortext[cor1]{Corresponding author}

\author[2]{Quanyong Zhu}

\affiliation[2]{organization={School of Mathematics and Computer Science, Lishui University},
            postcode={323000}, 
            state={Zhejiang},
            country={China}}
\author[1]{Ming Zhang}

\author[1]{Xiaobing Luo}

\author[3]{A. A. Shanenko}
\affiliation[3]{organization={HSE University}, 
				postcode={101000},
				city={Moscow},
				country={Russia}}

\begin{abstract}
Recently, a surface superconductor-insulator transition has been predicted for a bulk superconductor in an electric field applied perpendicular to its surface. The related calculations were performed within a one-dimensional Hubbard model by numerically solving the Bogoliubov-de Gennes (BdG) equations without the Hartree-Fock (HF) interaction potential. The phase diagram of the surface superconducting, metallic, and insulating states was obtained as dependent on the electric field and temperature. This diagram was found to be in agreement with experimental results reported previously for (Li,Fe)OHFeSe thin flakes. In the present work, by taking into account the HF potential, we find that the latter acts as a kind of an extra electrostatic potential that enhances the electric-field effects on the surface states. The qualitative features of the phase diagram remain the same but the surface superconductor-insulator transition occurs at significantly lower electric fields, which supports prospects of its experimental observation in bulk samples.
\end{abstract}


\begin{highlights}
\item The surface superconductor-metal-insulator transition induced by an applied electric field in a bulk superconductor is considered within a one-dimensional Hubbard model by numerically solving the Bogoliubov-de Gennes (BdG) equations.

\item By including the Hatree-Fock (HF) interaction in the BdG equations, it is demonstrated that the HF potential can be considered as a kind of an additional electrostatic potential enhancing the electric-field effects on the surface states.

\item Our study reveals that the critical electric fields of the surface supercon\-ductor-metal and surface metal-insulator transitions are significantly reduced when taking into account the HF interaction.
\end{highlights}


\end{frontmatter}


\section{Introduction}\label{sec1}

Effects of the electric field on the superconductor-metal transition have been investigated intensively since 1960s~\cite{Glover1960, Bonfiglioli1962, Meissner1967, Mannhart1991, Konsin1998,Xi1998, Szalowski2014, Bours2020, Amoretti2021, Amoretti2023}. In particular, by using Sn and In films as one of the condenser plates, researchers produced~\cite{Glover1960,Bonfiglioli1962} perpendicular electric fields leading to a slight shift of the film superconducting temperature about $\Delta T_c\approx \pm10^{-4}$~K. It was also reported~\cite{Staley2009} that for NbSe$_2$ such a shift can be as large as $0.2$ K ($\sim 8.0\%T_c$, with $T_c$ the critical superconducting temperature). Such results can be understood as follows. $T_c$ is connected with the single-particle density of states (DOS) at the Fermi level according to the BCS theory~\cite{Gennes1966}. When turning on an electric field, the single-particle spectrum changes and so does the DOS at the Fermi level. In addition, the charge-carrier density of thin films changes in the process of charging the film and hence, the Fermi level is altered. However, the latter mechanism makes much less contribution, as argued in the paper~\cite{Staley2009}. Thus, since the DOS changes, the electric-field-induced shift in $T_c$ occurs. For oxide superconductors, $T_c$ can be also modified by a sufficiently strong electric field due to the dielectric breakdown~\cite{Ahn1999,Ahn2003,Takahashi2004}. Besides $T_c$, even the critical supercurrent can be affected by the electric field~\cite{Golokolenov2021, Elalaily2021, Paolucci2021,Ritter2021,Amoretti2022}.

Moreover, an insulator-metal-superconductor transition can occur in the electric field of the dielectric breakdown of an insulator. In this scenario, a sufficiently strong electric field induces charge carriers in an insulator, which causes an attractive electron-electron interaction resulting in the superconducting order~\cite{Parendo2005,Ueno2008}. For example, being under the application of a gate voltage increasing from $0$ to $42.5$ V at $T=65$ mK, a $1$-nm-thick film of amorphous bismuth goes through an insulator-metal-superconductor transition since its resistance reduces from $22$ to $0$ k$\Omega$~\cite{Parendo2005}. Similar behavior has been found in SrTiO$_3$~\cite{Ueno2008,Paolucci2019}, $2$-nm-thick GdBa$_2$Cu$_3$O$_{7-x}$ films~\cite{Ahn2003}, atomically flat ZrNCl films~\cite{Ye2010}, La$_{2-x}$Sr$_x$CuO$_4$ films~\cite{Bollinger2011}, etc. 

Recently, a surface superconductor-insulator transition has been predicted for a bulk superconductor in a perpendicular electric field~\cite{Yin2023}. The consideration was done within a one-dimensional Hubbard model by numerically solving the Bogoliubov-de Gennes (BdG) equations. It was demonstrated that electrons are accumulated near (or removed from) the system edges due to the electric field applied parallel to the chain. Then, for sufficiently large fields, the sites near the chain edges become either fully occupied by electrons or completely empty, manifesting a surface insulating state. At zero temperature the surface superconductor-insulator transition occurs when increasing the field. At finite temperatures, the system exhibits the sequence of the surface superconductor-metal and surface metal-insulator transitions. Notice that these results are in a qualitative agreement with the phase diagram obtained by the transport measurements for (Li,Fe)OHFeSe thin films~\cite{Ma2019,Yin2020}. However, the BdG equations utilized in the paper~\cite{Yin2023} do not include the Hartree-Fock (HF) potential that is proportional to the election density~\cite{Gennes1966}. It is well-known that for a uniform spatial distribution of electrons, the effects of the HF interaction potential are nullified by the corresponding shift of the chemical potential. However, this is not the case for superconductors with a non-uniform electron density~\cite{Chen2009, Chen2014, Chen2016}. For example, the HF potential has a significant effect on the quantum-size oscillations of the critical temperature in nanoscale superconductors~\cite{Chen2009}. Thus, the question arises of how the HF self-consistent interaction affects the surface superconductor-insulator transition.

In the present work we investigate the effects of an external electric field on the surface states of a bulk superconductor within a one-dimensional Hubbard model by numerically solving the self-consistent BdG equations with the HF potential taken into account. Our numerical results demonstrate that the effect of the HF interaction potential on the surface states is similar to an additional electrostatic potential. As a result, the main qualitative features of the phase diagram displaying the switching between surface superconducting, metallic and insulating states, remain the same as compared to those without the HF interaction~\cite{Yin2023}. However, importantly, the critical electric field of the surface superconductor-insulator transition decreases significantly, almost by a factor of $2$.

The paper is organized as follows. Section~\ref{sec2} outlines the self-consistent BdG equations for a one-dimensional attractive Hubbard model in an external electric field. As is mentioned above, the HF interaction potential is included. In Sec.~\ref{sec3} we discuss our numerical results and the corresponding phase diagram of the surface superconducting, metallic and insulating states with and without the HF potential. Conclusions are given in Sec.~\ref{sec4}.

\section{Theoretical Formalism}\label{sec2}

Following Refs.~\cite{Yin2023,Bai2023,Tanaka2000}, we employ the Hubbard model of a one-dimensional chain of atoms with the grand-canonical Hamiltonian
\begin{equation}\label{hubbard}
\mathscr{H} - \mu \mathscr{N}_e  = -\sum_{i\delta\sigma}t_\delta c^{\dagger}_{i+\delta,\sigma}c_{i\sigma} + \sum_{i\sigma}\big[V(i) -\mu\big] n_{i\sigma} -g \sum_i n_{i\uparrow}n_{i\downarrow},
\end{equation}
where $c_{i\sigma}$ and $c^{\dagger}_{i\sigma}$ are the annihilation and creation operators of an electron with the spin projection $\sigma(=\uparrow,\downarrow)$ at sites $i=0,...,N+1$; $\mu$ and $g>0$ are the chemical potential and on-site attractive electron-electron interaction, respectively; $t_\delta$ is the hopping parameter for electrons between the sites $i$ and $i+\delta$~(we only consider the nearest neighbors, i.e., $\delta=\pm1$ and $t_\delta = t$); and $\mathscr{N}_e$ is the total electron number operator, i.e., $\mathscr{N}_e=\sum_{i\sigma} n_{i\sigma}$ and $n_{i\sigma}= c^{\dagger}_{i\sigma}c_{i\sigma}$). In addition, 
the on-site electrostatic potential energy $V(i)$~(for the electric field in the chain positive direction) is of the form~\cite{Yin2023}
\begin{equation}\label{V}
V(x) = -2\,q\lambda_E\,E_{0}\,e^{-L/2\lambda_{E}} \sinh\big[\big(2x-L\big)/2\lambda_{E}\big],
\end{equation}
where $q=-e$ is the electron charge, with $e$ being the elementary charge; $L=(N+1)a$ is the chain length, with $a$ the lattice constant; $x=(i-1)a$ is the site coordinate; $E_0$ is the strength of the screened electric field ${\bf E}(x)$; and $\lambda_{E}$ is the electric-field screening length. The corresponding screened electric field is given by 
\begin{equation}\label{Ei}
{\bf E}(x) = 2\,E_{0}\,e^{-L/2\lambda_{E}} \cosh\big[\big(2x-L\big)/2\lambda_{E}\big]\hat{\bf{x}},
\end{equation}
where $\hat{x}$ is the unit vector along the chain positive direction. Following Refs.~\cite{Yin2023,Bai2023}, we approximate $\lambda_{E}$ in the form $\lambda_E = \gamma\lambda_F$, where $\lambda_F$ is the Fermi wavelength taken at $E_0=0$. For the half-filling case, which is considered below, $\lambda_F = \sqrt{2}\pi a$~\cite{Yin2023}.

Utilizing the mean-field approximation~\cite{Gennes1966} for Eq.~(\ref{hubbard}), one obtains the effective Hamiltonian $H_{\rm eff}$~(for the s-wave pairing) in the form
\begin{align}\label{Heff}
H_{\rm eff} =& -t\sum_{i\delta\sigma} c^{\dagger}_{i+\delta,\sigma}c_{i\sigma} + \sum_{i\sigma}\big[V(i) + U_{\rm HF}(i) - \mu\big] n_{i\sigma}  \nonumber \\ 
& + \sum_i \big[\Delta(i)c^{\dagger}_{i\uparrow}c^{\dagger}_{i\downarrow} + \Delta^{*}(i)c_{i\downarrow}c_{i\uparrow}\big],
\end{align}
with $\Delta(i)$ the superconducting pair potential and $U_{\rm HF}(i)$ the Hartree-Fock single-electron interaction potential~\cite{Tanaka2000}. Diagonalizing $H_{\rm eff}$ with the Bogoliubov-Valatin transformation~\cite{Gennes1966}, we get the BdG equations~\cite{Bai2023, Tanaka2000}
\begin{subequations}\label{bdg}
\begin{align}
\epsilon_\alpha u_\alpha(i) & =  \sum_{i'} H_{ii'}u_\alpha(i') + U_{\rm HF}(i)u_\alpha(i) + \Delta(i) v_\alpha(i) \\
\epsilon_\alpha v_\alpha(i) & =  \Delta^*(i)u_\alpha(i)  - \sum_{i'} H^*_{ii'} v_\alpha(i') - U_{\rm HF}(i)v_\alpha(i),
\end{align}
\end{subequations}
where $H_{ii'}=-t\sum_{\delta=\pm1} \delta_{i',i+\delta}+\big[V(i)-\mu\big]\delta_{ii'}$; $\epsilon_\alpha$, $u_\alpha(i)$ and $v_\alpha(i)$ are the energy and wave functions of quasiparticles, respectively. The quantum number $\alpha$ enumerates the quasiparticle states in the energy ascending order. In our study the open boundary conditions are applied, that is, the quasiparticle wave functions vanish at the edge sites $i=0$ and $N+1$.

The chemical potential $\mu$ is determined by the electron-filling level $\bar{n}_e$ with the relations
\begin{equation}\label{ne}
\bar{n}_e = \frac{1}{N}\sum_i n_e(i), \quad n_e(i) = 2\sum_{\alpha}\big[ f_\alpha |u_\alpha(i)|^2 + (1-f_\alpha)|v_\alpha(i)|^2\big],
\end{equation}
where $n_e(i)$ is the averaged site occupation number (i.e. the electron spatial distribution) and $f_\alpha=f(\epsilon_\alpha)$ is the Fermi-Dirac distribution of bogolons. As is already mentioned above, in the present study we limit ourselves to consideration of the half filling, i.e., $\bar{n}_e=1$. Our conclusions are not sensitive to this choice, and other variants of $\bar{n}_e$ produce similar qualitative conclusions.

The pair potential $\Delta(i)$ and the HF single-electron potential $U_{\rm HF}(i)$ are determined by the quasiparticle energies and wave functions~\cite{Gennes1966, Bai2023, Chen2022}
\begin{align}
\Delta(i) &= g\sum_\alpha u_\alpha(i) v^*_\alpha(i)\big[1-2f_\alpha \big], \label{op} \\
U_{\rm HF}(i) & = -g\sum_\alpha \big[ |u_\alpha(i)|^2f_\alpha + |v_\alpha(i)|^2(1-f_\alpha) \big]. \label{HF}
\end{align}
The summation in Eq.~(\ref{op}) is over the quasiparticle species with the positive energies in the Debye window around the Fermi level, i.e. $0\leq\epsilon_\alpha \leq \hbar\omega_D$, with $\omega_D$ the Debye frequency. The summation in Eq.~(\ref{HF}) includes all the positive-energy quasiparticle species.~\cite{Tanaka2000}

The self-consistent calculation procedure is as follows. First, we solve the BdG Eqs.~(\ref{bdg}) using some initial guess for $\mu$, $\Delta(i)$, and $U_{\rm HF}(i)$. Second, using the obtained quasiparticle energies and wave functions, we calculate $n_e(i)$ together with new distributions $\Delta(i)$ and $U_{\rm HF}(i)$ from Eqs.~(\ref{ne})-(\ref{HF}). Third, we adjust $\mu$ to fit the half-filling regime using Eq.~(\ref{ne})~[the quasiparticle energies and wave functions are not altered in this step]. Then, these three steps are repeated until the convergence of the whole procedure. 

Below the energy-related quantities [e.g., $\Delta(i)$, $U_{\rm HF}(i)$, $\mu$, $V(i)$, and $g$], the length quantities (e.g., $\lambda_E$, $\lambda_F$, and $L$), and the edge electric field $E_0$ are given in units of the hopping parameter $t$, the lattice constant $a$, and the ratio $t/ea$, respectively. We set $\gamma=2$, $N=301$, $\hbar\omega_D=10$, and $g=2$, which are the same as in Ref.~\cite{Yin2023}. It is of importance to note that the qualitative conclusions of our study are not sensitive to the particular choice of the model parameters.

\section{Results and Discussions}\label{sec3}
\begin{figure}[htpb]
\centering
\includegraphics[width=0.65\textwidth]{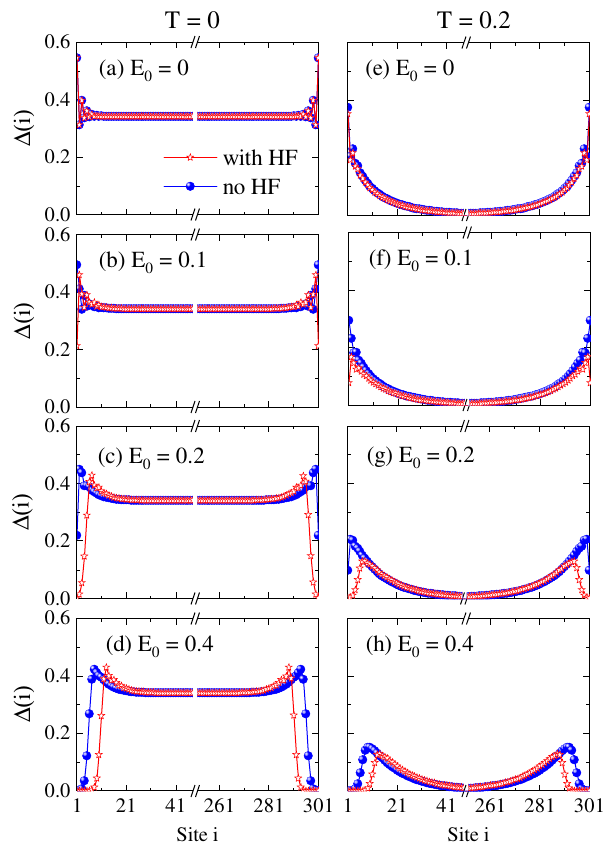}
\caption{The superconducting pair potential $\Delta(i)$ versus the site number $i$ at $T=0$ and $0.2$ for the electric-field strengths $E_0=0$, $0.1$, $0.2$ and $0.4$. The data marked by red stars are the results from the BdG equations with the HF potential while blue spheres represent the case without the HF interaction potential. $\Delta(i)$, $T$ and $E_0$ are given in units of $t$, $t/k_B$ and $t/ea$, respectively, with $e$ the absolute value of electron charge and $a$ the lattice constant.}
\label{fig1}
\end{figure}

\begin{figure}[htpb]
\centering
\includegraphics[width=0.65\textwidth]{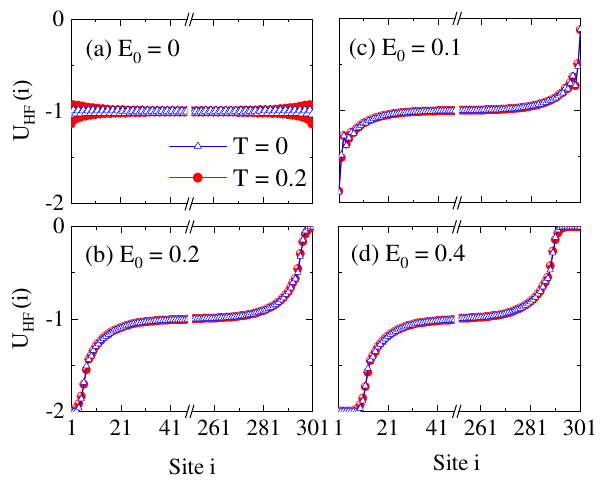}
\caption{The HF potential $U_{\rm HF}(i)$ (in units of $t$) for the same parameters as in Fig.~\ref{fig1}. Blue triangles correspond to $T=0$ while red spheres are for $T=0.2$.}
\label{fig2}
\end{figure}

Figure~\ref{fig1} illustrates the pair potential $\Delta(i)$ obtained with (red stars) and without (blue spheres) the HF potential for $E_0=0$, $0.1$, $0.2$ and $0.4$ at $T=0$ (a-d) and $0.2$ (e-h). The second temperature value is larger than the bulk critical temperature but smaller than the surface superconducting temperature (for more detail, see Refs.~\cite{Yin2023,Bai2023}). One can see that the profiles of $\Delta(i)$ are symmetric with respect to the center of the chain $i=151$. The corresponding HF potential is shown versus the site number $i$ in Fig.~\ref{fig2}. From Fig.~\ref{fig1}(a), it can be seen that the results with and without the HF interaction potential are the same throughout the entire chain in the absence of an electric field and at zero temperature. This is because the HF potential is spatially uniform at $T=0$ and $E_0=0$, we have $U_{\rm HF}(i)=-1$ [see the blue triangles in Fig.~\ref{fig2}(a)]. Thus, in this case, the only effect of including $U_{\rm HF}(i)$ is reduced to shifting the chemical potential that decreases from $0$ to $-1$. Recall that such a shift negates the effect of the appearance of the HF potential in the single-electron spectrum for uniform superconducting condensates.  

Taking $T=0.2$ and $E_0=0$, as illustrated in Fig.~\ref{fig1}(e), one finds that the values of $\Delta(i)$ calculated from the BdG equations with the HF potential, are slightly smaller than those for the model with the HF interaction near the chain edges. Such a deviation is connected with minor oscillations of $U_{\rm HF}(i)$ for $i < 21$ and $i>281$ [see the red spheres in Fig.~\ref{fig2}(a)]. At the same time, there is no difference between the pair potentials calculated with and without the HF contribution sufficiently far from the edges (deep in the chain), where $U_{\rm HF}(i)$ is nearly constant. 

For $E_0=0.1$ and $T=0$, see Fig.~\ref{fig1}(b), $\Delta(i=1)$ drops from $0.49$ obtained with the HF potential to $0.21$ calculated without the HF contribution. For $E_0=0.1$ and $T=0.2$, see Fig.~\ref{fig1}(f), $\Delta(i=1)$ drops from $0.29$~(without HF) to $0.07$~(with HF). In both cases, $\Delta(i)$ far from the edges is not affected, no matter whether the HF potential is taken into account or not. Similar results can be seen in Figs.~\ref{fig1}(c,g) corresponding to $E_0=0.2$~(for $T=0$ and $0.2$). However, for $E_0=0.4$, see Figs.~\ref{fig1}(d,h), $\Delta(i=1)$ is zero for both the calculations with and without the HF potential. Here, to observe the decrease of the pair potential due to including the HF interaction, one should choose $i\approx5-10$ rather than $i=1$. Indeed, the pair potential drops to zero near the chain edges for sufficiently large electric fields, which is the reflection of the surface superconductor-insulator transition at $T=0$ and the sequence of the surface superconductor-metal and surface metal-insulator transitions at finite temperatures, see Ref.~\cite{Yin2023}.

One can learn from Fig~\ref{fig1} that the HF interaction extends the suppression region of the superconducting condensate near the edges. This effect is directly connected with the fast changes in $U_{HF}(i)$ near the both edges for $E_0 > 0$, see Fig.~\ref{fig2}(c-d). Notice that $U_{HF}(i)$ is not very sensitive to the temperature and proportional to the electron spatial distribution $n_e(i)$, as is seen from Eqs.~(\ref{ne}) and (\ref{HF}). Thus, we arrive at the conclusion that when the external electric field suppresses the superconducting condensate near the chain edges, the HF potential enhances this effect, acting as a kind of an additional electrostatic potential. This is also seen from the comparison of the spatial profiles of $U_{HF}(i)$ and $V(i)$, the latter is discussed in Ref.~\cite{Yin2023}.   

\begin{figure}[htpb]
\centering
\includegraphics[width=0.65\textwidth]{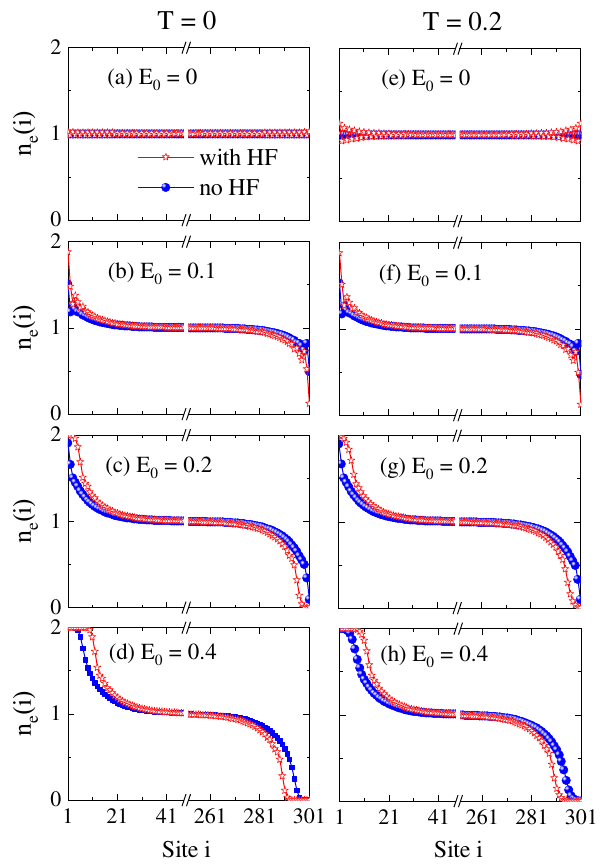}
\caption{The electron spatial distribution $n_e(i)$ versus the site number $i$ for the electric-field strength $E_0=0$, $0.1$, $0.2$ and $0.4$ at $T=0$ and $0.2$. The results of the BdG equations with the HF interaction are given by the red stars while those corresponding to the BdG equations without the HF potential are shown by blue solid spheres.}
\label{fig3}
\end{figure}

To proceed further, we study how the HF potential affects the spatial electron distribution $n_e(i)$. The results for $n_e(i)$ are used to specify the surface states near the edges together with those for $\Delta(i)$, see Ref.~\cite{Yin2023}. Figure~\ref{fig3} shows $n_e(i)$ calculated from the BdG equations with and without the HF contribution for $E_0=0$, $0.1$, $0.2$ and $0.4$ at $T=0$ and $0.2$. The results with the HF potential are marked by red stars while those without the HF potential are labeled by blue spheres. According to Eqs.~(\ref{ne}) and (\ref{HF}), one finds $U_{\rm HF}(i)=-(g/2)n_e(i)$. This is the reason why the spatial profiles of $U_{\rm HF}(i)$ given in Fig.~\ref{fig2} are similar to the spatial distributions of $n_e(i)$ in Fig.~\ref{fig3}. 

In the absence of the electric field, $n_e(i)$ calculated without the HF interaction at $T=0$ and $0.2$ is almost uniform [see the curves with blue spheres in Figs.~\ref{fig3}(a) and (e)]. When including the HF potential at $E_0=0$, $n_e(i)$ is also uniform for $T=0$~[see the curve with red stars in Fig.~\ref{fig3}(a)] while it exhibits weak oscillations near the chain edges at $T=0.2$~[see red starred curves in Fig.~\ref{fig3}(e) and compare with Fig.~\ref{fig2}(a)]. 

\begin{figure}[htpb]
\centering
\includegraphics[width=0.7\textwidth]{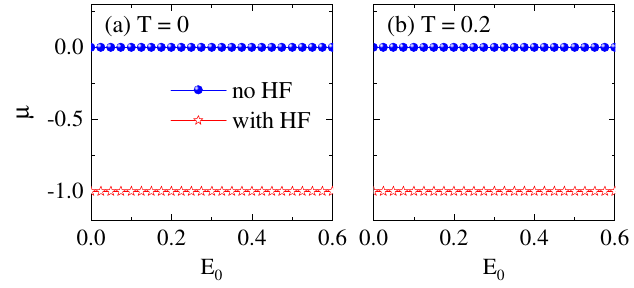}
\caption{The chemical potential $\mu$ (in units of $t$) as a function of the electric-field strength $E_0$ at $T=0$~(a) and $0.2$~(b). The data obtained from the BdG equations without the HF contribution are labeled by blue spheres while those with the HF potential are marked by red stars.}\label{fig4}
\end{figure}

For $E_0=0.1$ both at $T=0$ and $0.2$, see panels (b) and (f), $n_e(i=1)$ obtained with the HF potential is larger while $n_e(i=301)$ is smaller than their counterparts calculated without the HF contribution. The same conclusion holds also for Figs.~\ref{fig3}(c,g). However, for Figs.~\ref{fig3}(d,h) similar results are obtained for the sites with $i\approx 5-10$ while $n_e(i=1)=2$ and $n_e(i=301)=0$ for the both cases with and without the HF potential. This is the clear signature of the surface insulating states, see Ref.~\cite{Yin2023}. The accumulation of electrons at the left edge and their depletion at the left edge are the reasons for the formation of the surface insulator. One can see that the surface insulating state for the system with the HF interaction occurs at $E_0\approx 0.2$ in Fig.~\ref{fig4}(c). However, without the HF contribution we get the zero-temperature critical field $E_0^*=0.35$, see Fig.~1(f) of Ref.~\cite{Yin2023}. Overall, we again conclude that the HF potential enhances the electric-field effects on the surface states, acting similarly to an additional electrostatic potential, which shifts the surface superconductor-insulator transition (for $T=0$) and the sequence of the superconductor-metal and metal-insulator transition (for $T > 0$) to lower electric fields.

Now we investigate how the chemical potential $\mu$ is influenced by including the HF potential $U_{HF}(i)$. $\mu$ satisfies Eq.~(\ref{ne})~[with $\bar{n}_e=1$ at the half filling] in which the wave functions and energies of quasiparticles are affected by $U_{\rm HF}(i)$ and $E_0$ through Eqs.~(\ref{bdg}). In Fig.~\ref{fig4}, the values of $\mu$ with (red stars) and without (blue spheres) the HF interaction taken into account, are given as a function of $E_0$ at $T=0$~(a) and $0.2$~(b). Surprisingly, from Fig.~\ref{fig4}(a) we find that at $T=0$, the both values of $\mu$ are nearly constant when varying $E_0$: $\mu\approx0$ for the case without $U_{\rm HF}(i)$ while $\mu\approx-1$ in the opposite variant. Moreover, the behavior of $\mu$ at $T=0.2$ is nearly the same as that for $T=0$. 

These results for $\mu$ can be understood as follows. The single-electron energy reads $\xi_k=-2t{\rm cos}(ka)$ for $g=0$, $E_0=0$ and $U_{\rm HF}(i)=0$ and, so, the half-filling condition leads to $\mu=0$, i.e., the chemical potential is located at the center of the single-electron energy band and this values is almost not sensitive to the change of the temperature from $0$ to $0.2$. When turning on the electron-electron interaction with $g=2$, $\mu$ without the HF potential decreases from $0$ to $-1.2\times10^{-5}$ at $T=0$ while for $T=0.2$ it is reduced to $-0.3\times10^{-5}$. Hence, $\mu$ remains very close to $0$, see the blue spheres in Figs.~\ref{fig4}(a) and (b). When including the HF interaction, $U_{\rm HF}(i)$ is nearly symmetric in energy about the value $\sim-1$ for all $E_0$ because $n_e(i)$ is almost symmetric relative to the value $\bar{n}_e=1$. As a consequence, the center of the single-particle energy band with the HF potential is shifted to the value $\approx -1$ and so does the chemical potential in the half-filling regime, see the red stars in Figs.~\ref{fig4}(a) and (b).

\begin{figure}[htpb]
\centering
\includegraphics[width=0.55\textwidth]{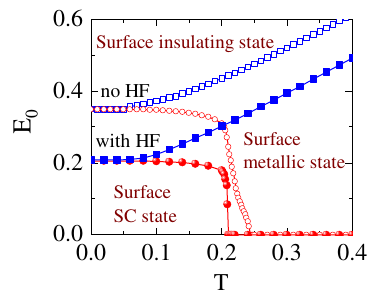}
\caption{The phase diagram of the surface superconducting (SC), metallic and insulating states in the $E_0$-$T$ plane. The boundary curves marked by solid symbols are the results with the HF potential, while those with open symbols correspond to the data calculated without the HF interaction~\cite{Yin2023} and shown as a reference.}\label{fig5}
\end{figure}

Finally, the phase diagram of the surface states obtained from the BdG equations with the HF potential is given in $T$-$E_0$ plain in Fig.~\ref{fig5}, where its boundaries are labeled by solid symbols. For comparison, the phase diagram corresponding to the BdG equations without the HF contribution~\cite{Yin2023} is also shown in Fig.~\ref{fig5} , and its boundaries are marked by open symbols. The curve with the red spheres separates the surface superconducting (SC) and surface metallic state while the curve with the blue squares is the boundary between the surface insulating and surface metallic states. Following Ref.~\cite{Yin2023}, we consider that for the surface SC state one has $\Delta(i=1) > 0$ and $n_e(i=1) \neq 2$; for the surface metallic state $\Delta(i=1) = 0$ and $n_e(i=1) \neq 2$; and, finally, for the surface insulating state $\Delta(i=1)=0$ and $n_e(i=1) = 2$.

One can find from Fig.~\ref{fig5} that qualitatively, the phase diagram of the three surface states is same for the both cases, with and without the HF potential. However, the direct surface superconductor-insulator transition for the model with the HF potential occurs at zero temperature at the lower field-strength $E_0=E_0^*=0.20$ as compared to $E_0=E_0^*=0.35$ of the case without the HF interaction. One finds that $E^*_0$ is reduced significantly, by almost a factor of $2$~(by $43\%$). One can also see that the temperature-dependent critical electric fields of both the surface superconductor-metal and surface metal-insulator transitions are also lowered in the case with the HF potential. According to our discussions of the results in Figs.~\ref{fig1} and \ref{fig3}, this significant reduction of the critical fields is connected with the fact that the HF interaction enhances the electric-field effects on the surface states, acting as a kind of an additional electrostatic potential.

It has been predicted in Ref.~\cite{Yin2023}~(without the HF interaction) that the direct electric-field-induced surface superconductor-insulator transition was possibly observed in SrTiO$_3$ films~\cite{Wang2001} at a critical field lower than the dielectric-breakdown field. Our present results of the self-consistent BdG equations with the HF potential suggest that such a transition can occur at significantly lower fields.

\section{Conclusions}\label{sec4}

In conclusion, we have investigated the effect of including the HF potential in the BdG equations on the electric-field-induced surface superconductor-insulator transition reported in Ref.~\cite{Yin2023}. Our study is based on a one-dimensional attractive Hubbard model at the half filling. Our study reveals that including the HF interaction between electrons enhances the electric-field effects on the surface states. The HF potential can be considered as a kind of an additional electrostatic potential so that the critical electric fields of the superconductor-metal and metal-insulator transitions significantly decrease in the presence of the HF interaction, as compared to those without the HF potential. The qualitative features of the phase diagram of the surface superconducting, metallic, insulating states remain the same.

\section*{CRediT authorship contribution statement}
All authors certify that they have participated sufficiently in the work to take public responsibility for the content, including participation in the concept, writing, or revision of the manuscript. 

\section*{Declaration of competing interest}
The authors declare that they have no known competing financial interests or personal relationships that could have appeared to influence the work reported in this paper.

\section*{Acknowledgements}
This work was supported by Science Foundation of Zhejiang Sci-Tech University(ZSTU) (Grants No. 19062463-Y \& 22062336-Y), Open Foundation of Key Laboratory of Optical Field Manipulation of Zhejiang Province (ZJOFM-2020-007). The study has also been funded within the framework of the HSE University Basic Research Program.





\end{document}